\documentclass[twocolumn]{aastex701}
\usepackage{graphicx}
\usepackage{float}
\usepackage{amsmath}
\usepackage{amssymb}
\usepackage{subfigure}
\usepackage{xcolor}
\usepackage{booktabs}
\usepackage[flushleft]{threeparttable}
\usepackage{multirow}
\usepackage{bm}

\hypersetup{
   colorlinks,
   linkcolor={blue!88!black!80},
   citecolor={blue!88!black!80},
   urlcolor={blue!88!black!80}}
\begin{document}

\title{The Intermediate-Mass Black Hole Reverberation Mapping Project:\\Stable Optical Continuum Lags of an IMBH in the Dwarf Galaxy NGC 4395 Over Years}

\correspondingauthor{ Hengxiao Guo, Xiaowei Liu}
\email{hengxiaoguo@gmail.com \\ x.liu@ynu.edu.cn} 

\author[orcid=0009-0002-7625-2653]{Yu Pan}
\affiliation{South-Western Institute for Astronomy Research, Yunnan Key Laboratory of Survey Science, Yunnan University, Kunming, Yunnan 650500, People's Republic of China}
\email{yupan0304@163.com}

\author[orcid=0000-0001-8416-7059]{Hengxiao Guo}
\affiliation{Shanghai Astronomical Observatory, Chinese Academy of Sciences, 80 Nandan Road, Shanghai 200030, People's Republic of China}
\email{hengxiaoguo@gmail.com}

\author[orcid=0000-0001-5561-2010]{Chenxu Liu}
\affiliation{South-Western Institute for Astronomy Research, Yunnan Key Laboratory of Survey Science, Yunnan University, Kunming, Yunnan 650500, People's Republic of China}
\email{cxliu@ynu.edu.cn}

\author[orcid=0009-0000-4068-1320]{Xinlei Chen}
\affiliation{South-Western Institute for Astronomy Research, Yunnan Key Laboratory of Survey Science, Yunnan University, Kunming, Yunnan 650500, People's Republic of China}
\email{xlchen@stu.ynu.edu.cn}

\author[orcid=0009-0006-1010-1325]{Yuan Fang}
\affiliation{South-Western Institute for Astronomy Research, Yunnan Key Laboratory of Survey Science, Yunnan University, Kunming, Yunnan 650500, People's Republic of China}
\email{fangyuan@ynu.edu.cn}

\author[orcid=0000-0002-2510-6931]{Jinghua Zhang}
\affiliation{South-Western Institute for Astronomy Research, Yunnan Key Laboratory of Survey Science, Yunnan University, Kunming, Yunnan 650500, People's Republic of China}
\email{zhang_jh@ynu.edu.cn}

\author[0000-0002-4521-6281]{Wenwen Zuo}
\affiliation{Shanghai Astronomical Observatory, Chinese Academy of Sciences, 80 Nandan Road, Shanghai 200030, People's Republic of China} 
\email{wenwenzuo@shao.ac.cn}

\author[0000-0002-8186-4753]{Philip G. Edwards} 
\affiliation{CSIRO Space and Astronomy, PO Box 76, Epping, NSW, 1710, Australia}
\email{philip.Edwards@csiro.au}

\author[0000-0002-5841-3348]{Jamie Stevens}
\affiliation{CSIRO Space and Astronomy, PO Box 76, Epping, NSW, 1710, Australia}
\email{jamie.Stevens@csiro.au}

\author[0009-0008-7583-5658]{Manqi Fu}
\affiliation{Department of Astronomy, Xiamen University, Xiamen, Fujian 361005, People's Republic of China} 
\email{fumanqi@stu.xmu.edu.cn}

\author[0000-0002-0771-2153]{Mouyuan Sun}
\affiliation{Department of Astronomy, Xiamen University, Xiamen, Fujian 361005, People's Republic of China} 
\email{msun88@xmu.edu.cn}

\author[0000-0002-4223-2198]{Zhen-yi Cai}
\affiliation{Department of Astronomy, University of Science and Technology of China, Hefei, Anhui 230026, People's Republic of China} 
\affiliation{School of Astronomy and Space Science, University of Science and Technology of China, Hefei 230026, People's Republic of China} 
\email{zcai@ustc.edu.cn}

\author[orcid=0000-0002-8109-7152]{Guowang Du}
\affiliation{South-Western Institute for Astronomy Research, Yunnan Key Laboratory of Survey Science, Yunnan University, Kunming, Yunnan 650500, People's Republic of China}
\email{dugking@ynu.edu.cn}

\author[orcid=0009-0006-5847-9271]{Xingzhu Zou}
\affiliation{South-Western Institute for Astronomy Research, Yunnan Key Laboratory of Survey Science, Yunnan University, Kunming, Yunnan 650500, People's Republic of China}
\email{zxz_zoe@126.com}

\author[orcid=0009-0005-8762-0871]{Tao Wang}
\affiliation{South-Western Institute for Astronomy Research, Yunnan Key Laboratory of Survey Science, Yunnan University, Kunming, Yunnan 650500, People's Republic of China} 
\email{wangtao@itc.ynu.edu.cn}

\author[orcid=0009-0003-6936-7548]{Xufeng Zhu}
\affiliation{South-Western Institute for Astronomy Research, Yunnan Key Laboratory of Survey Science, Yunnan University, Kunming, Yunnan 650500, People's Republic of China}
\email{xufengzhu@ynu.edu.cn}

\author[orcid=0000-0003-0394-1298]{Xiangkun Liu}
\affiliation{South-Western Institute for Astronomy Research, Yunnan Key Laboratory of Survey Science, Yunnan University, Kunming, Yunnan 650500, People's Republic of China}
\email{liuxk@ynu.edu.cn}

\author[orcid=0000-0003-1295-2909]{Xiaowei Liu}
\affiliation{South-Western Institute for Astronomy Research, Yunnan Key Laboratory of Survey Science, Yunnan University, Kunming, Yunnan 650500, People's Republic of China} 
\email{x.liu@ynu.edu.cn}

\begin{abstract}
NGC 4395 is a nearby dwarf spiral galaxy hosting an active galactic nucleus (AGN) powered by an intermediate-mass black hole (IMBH, $M_{\rm BH} \sim 10^{4-5}\,M_\odot$). Recent optical continuum reverberation mapping studies have suggested potential lag variations between different epochs, offering important clues to the physical mechanisms governing variability in the vicinity of the central black hole. We present continuous intranight multi-band photometric monitoring of NGC 4395 based on five nights of observations, including three nights from the Faulkes Telescope North (two archival) and two new nights from Mephisto. This represents the first systematic investigation of optical continuum lag stability in a robustly confirmed IMBH. By applying difference-imaging techniques to both the new observations and the reprocessed archival data, we significantly detect optical inter-band lags of $\sim$5--15 minutes, which increase monotonically with wavelength. No obvious $u$-band lag excess is observed, implying a negligible fractional contribution from diffuse continuum (DC) emission to the optical continuum, in agreement with our spectral decomposition results. Remarkably, the inter-band lags remain stable over multi-year baselines. We suggest that this long-term lag stability may be related to the minor DC contribution, a relatively steady disk–corona structure, and the unusually high X-ray–to–optical luminosity ratio characteristic of low-luminosity AGNs, which likely allows X-ray reprocessing to dominate over other potential variability mechanisms. Future facilities like Gemini/SCORPIO, with its simultaneous optical-to-near-infrared coverage, will be ideally suited to play an important role in advancing this field.

\end{abstract}
\keywords{\uat{Active galactic nuclei}{16} --- \uat{Intermediate-mass black holes}{816} --- \uat{Accretion disks}{43} --- \uat{Reverberation mapping}{2019} --- \uat{NGC 4395}{1108}}

\section{Introduction}\label{sec_intro}
Continuum reverberation mapping (CRM) is a powerful observational tool for probing the structure of accretion disks in active galactic nuclei (AGNs) \citep[for a review, see][]{Cackett21}. Within the framework of the standard optically thick, geometrically thin accretion disk \citep[SSD,][]{Shakura73}, X-ray reprocessing (i.e., the lamp-post model), in which a compact hard X-ray–emitting corona illuminates the accretion disk and is reprocessed into lower-energy UV/optical emission, predicts a characteristic lag–wavelength relation of $\tau \propto \lambda^{4/3}$ \citep{Krolik91,Cackett07}. Early intensive monitoring of nearby AGNs provided observational support for this relation, albeit with relatively low significance \citep[e.g.,][]{Collier98,Sergeev05}. Subsequent progress in this field has been largely hindered by observational limitations, particularly incomplete wavelength coverage and insufficient temporal sampling, which restrict precise measurements of short-term continuum lags.

A major leap forward was achieved by the AGN STORM campaign on NGC~5548, which delivered high-cadence, multi-wavelength, long-term monitoring with unprecedented temporal coverage from X-ray to optical bands using multiple ground- and space-based telescopes \citep[e.g.,][]{DeRosa15,Fausnaugh16}. The exceptional data revealed strongly correlated variability across the UV and optical bands, while the correlation between X-ray and UV variations was found to be significantly weaker \citep{Edelson15}. These observations established that continuum lags increase with wavelength. 
Unexpectedly, the observation revealed that the inferred accretion disk size is systematically larger, by a factor of several, than predicted by the SSD model, raising the so-called ``disk size crisis''
\citep{McHardy14,Edelson15,Fausnaugh16}, which is also independently revealed by microlensing studies \citep[e.g.,][]{Morgan10}. Subsequent studies have generally confirmed these results in individual sources \citep{Cackett18,Edelson19,Kara21,Vincentelli21,Lewin24,Liu24,Panagiotou25} and through sample analyses based on multi-season light curves from time-domain surveys \citep{Jiang17,Mudd18,Homayouni19,Yu20b,Guo22,Guo22b,Jha22}.

A particularly noteworthy result from studies of NGC~5548 is the detection of a pronounced lag excess in the $u/U$-band, manifested as a discontinuity around the Balmer break at $\sim$3646~\AA\ in vacuum. This feature is clearly evident in the high-quality lag spectrum of NGC~4593 \citep{Cackett18} and NGC~4151 \citep{Feng25}, and similar behavior has been reported in other AGNs \citep{Edelson19,Kammoun21b,Kara23,Lewin23,Lewin24}, although it is not observed in every object \citep{Kara21,Wang23}. These results are reminiscent of a contribution from diffuse continuum (DC) emission, given the similarity between the observed lag spectrum and the predicted DC flux density spectrum \citep{Korista01}. The DC component arises primarily from free--bound and free--free emission of the inner broad-line region (BLR) gas, with a characteristic emitting scale suggested to be about half of the H$\beta$-emitting size \citep{Netzer22}. Taking into account its larger emitting size, it may help resolve or alleviate part of the accretion-disk size discrepancy inferred from CRM \citep[e.g.,][]{Lawther18,Chelouche19,Korista19,Netzer22,Guo22a,Jaiswal23}.

Proposed explanations for the unusually large continuum lags in AGNs can be broadly divided into two classes. One class modifies the underlying accretion-disk theory or the X-ray reprocessing scenario to increase the predicted continuum lags, for example through altered temperature–radius relations, as expected in disk-wind–modified disks \citep[e.g.,][]{Sun19,Li19}, non-blackbody emission from the accretion disk \citep{Hall18}, more realistic relativistic X-ray reprocessing \citep{Kammoun19}, or intrinsically fluctuating accretion disks driven by local temperature variations, either without X-ray reprocessing \citep[e.g.,][]{Cai18,Cai20} or combined with X-ray irradiation \citep{Sun20}. The other class invokes additional emission components to reconcile the discrepancy between the theoretically predicted and observed continuum lags, such as diffuse continuum emission from the broad-line region \citep[e.g.,][]{Gardner17,Chelouche19}, disk winds \citep{Netzer22}, or extended UV reprocessors \citep[e.g.,][]{Jaiswal23,Chan25}. Whether X-ray reprocessing is efficient, or how these mechanisms jointly shape the observed lags, remains an open question.

\begin{figure*}
\centering
  \includegraphics[width=1.\textwidth]{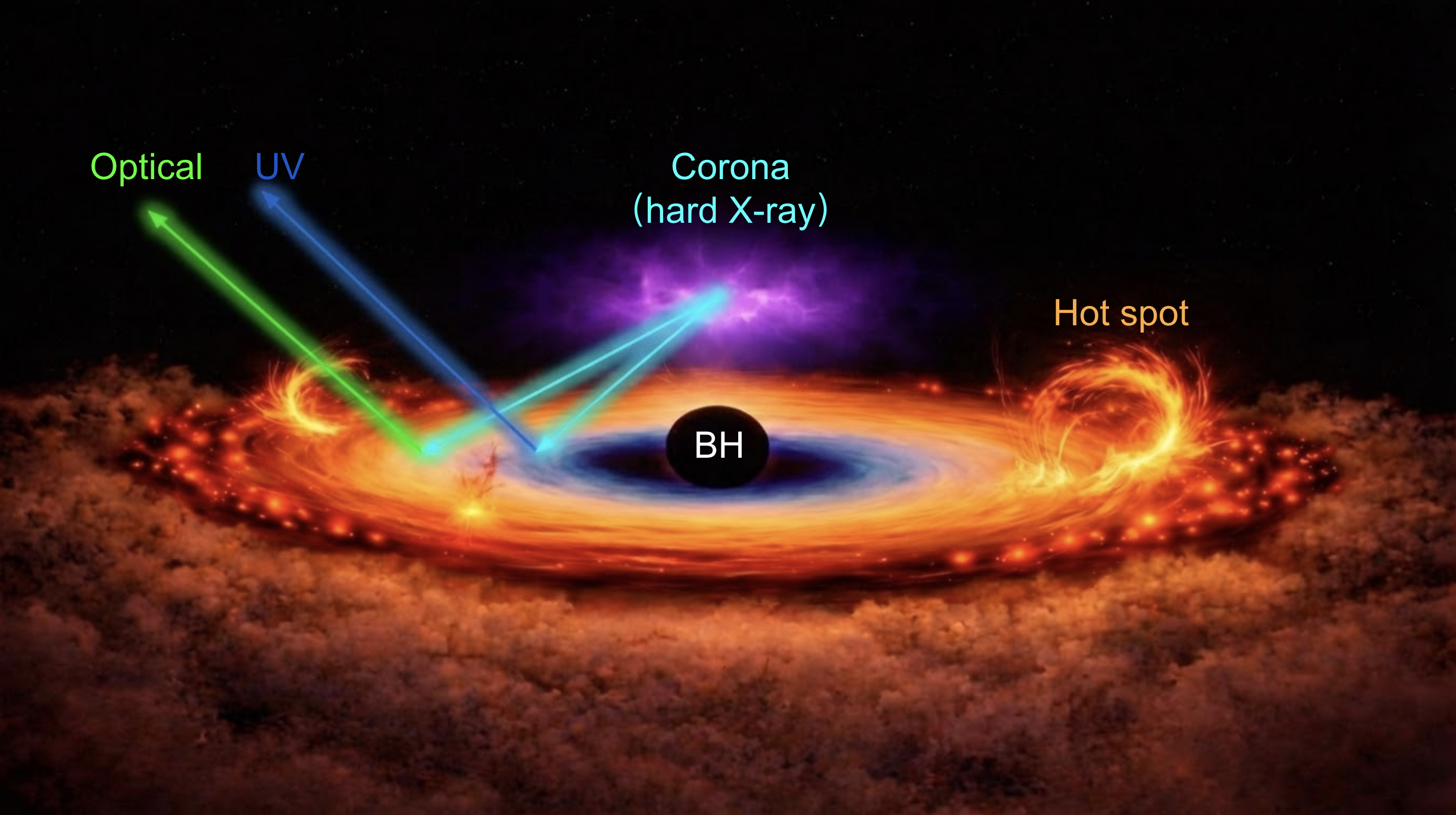}
  \caption{Schematic illustration of the black hole–accretion disk system and the continuum reverberation mapping (CRM) process. Variable hard X-rays from a compact corona irradiate an optically thick, geometrically thin accretion disk, producing wavelength-dependent continuum emission via thermal reprocessing. The arrows denote the light paths and the associated light-travel–time delays at different wavelengths. Intrinsic disk perturbations, such as hot spots, are also illustrated.
  The cartoon illustration was originally designed by us and later enhanced using Gemini
Nano Banana to improve its clarity and visual presentation.}
  \label{fig:crm}
\end{figure*}

Figure \ref{fig:crm} presents a schematic view of the classical X-ray reprocessing scenario, together with intrinsic disk fluctuations, such as hot spots driven by magnetically induced turbulence or reconnection and local temperature fluctuations.

Although the physical origin of continuum lags remains under debate, exploring their temporal variability may provide crucial clues for distinguishing among different mechanisms \citep{Su24b}. Here, ``optical continuum lag variability'' refers to changes in the measured inter-band continuum lags between different observing epochs. Such variability has been reported in several AGNs, including NGC~7469 \citep{Vincentelli23}, Fairall~9 \citep{Pal17,HernándezSantisteban20,Edelson24}, and NGC~4151 \citep{Zhou25}. Moreover, a systematic study of $\sim$100 AGNs by \citet{Su25} found that roughly half of the sources exhibit significant lag variations, suggesting that such variability may be common. However, disentangling whether the observed lag variations arise from observational issues or from genuinely intrinsic physical processes remains highly challenging and time-consuming.

Given the low black hole masses and luminosities, intermediate-mass black holes (IMBHs) such as NGC~4395 exhibit intrinsically short UV/optical variability timescales and minute-level inter-band lags \citep{Peterson05,Desroches06,McHardy16,Woo19}. These properties make IMBHs a unique laboratory for high-cadence and repeated measurements of continuum-lag variability. Using two nights of high-cadence optical monitoring, \citet{Montano22} (hereafter MJ22) found that the continuum lags differed significantly between the two nights, particularly at longer wavelengths, with the second night exhibiting a turnover to a flatter lag–wavelength relation that deviates from the predictions of the SSD model. Similarly, \citet{McHardy23} (hereafter MI23) reported a comparable turnover trend in a single-night campaign, yet with overall smaller lags, further highlighting the potential lag variation and the complexity of the underlying variability mechanisms.

Therefore, in this work, we present a systematic multi-band CRM study of NGC 4395 spanning several years. This includes both new observations and a reanalysis of previously published data using improved difference-imaging photometry techniques. We assess the temporal stability of inter-band continuum lags and explore the implications for accretion-disk structure and variability mechanisms in IMBHs. This work is the second paper of our IMBH-RM project (Guo H. in prep.), and more details can be found in \citet{Sun25}. The paper is organized as follows. Section~\ref{sec_obs} describes the observations and data sets. Section~\ref{sec_reduction} details the extraction of multi-band light curves and the measurement of inter-band time lags. Section~\ref{sec_result and discussion} presents our main results and their interpretation. Finally, Section~\ref{sec_summary} summarizes our conclusions and discusses future prospects.

\section{Observations}\label{sec_obs}
NGC~4395 is located at a distance of 4.6~Mpc based on TRGB measurements \citep{Karachentsev13}. It hosts a well-established IMBH with an estimated mass of $M_{\rm BH} \sim 2\times10^{4}$--$4\times10^{5}\,M_\odot$ based on gas-dynamical and reverberation-mapping measurements \citep{Peterson05,Brok15,Woo19}. The SED-based bolometric luminosity is $1.52\times10^{41}$\,erg~s$^{-1}$, yielding an average Eddington ratio of $\sim$0.06 when adopting the mean black hole mass within the above range \citep{Kammoun19}. Previous observations have shown ubiquitous intra-night optical variability at $\sim$0.1~mag, making this source well suited for CRM. We previously obtained two nights of high-cadence monitoring with LCOGT/FTN (MJ22), detecting correlated variability across the $griz$ bands and interband lags of $\sim$8--21 minutes relative to the $g$ band using the full light curves. When analyzed separately, the second night shows weaker cross-correlation amplitudes and systematically shorter lags than the first. In this work, we add one additional night of FTN observations and two new nights from Mephisto. We reanalyze all five nights using difference-imaging techniques to better isolate nuclear variability and assess the robustness of the lag measurements.

\subsection{LCOGT/FTN} \label{sec_lco}

We observed NGC~4395 on 2022 April 26--27 and 2025 March 26 (UT) with MuSCAT3 \citep{Norio20} on the 2~m Faulkes Telescope North (FTN) at Haleakalā Observatory, a node of the Las Cumbres Observatory Global Telescope (LCOGT) Network \citep{Brown13}. 
The observations consist of continuous intra-night monitoring with a total duration of $\sim$6--8~hr per night. MuSCAT3 provides simultaneous four-band imaging in $g', r'$, $i'$, and $z_s$ with a $9\farcm1\times9\farcm1$ field of view and a pixel scale of $0\farcs27$. We used exposure times of 100~s in $g'$, $i'$, and $z_s$. The exposure time in $r'$ was 25~s in 2022 and 100~s in 2025, yielding a total of $\sim$200--300 epochs per band. Observing conditions were generally clear over the three nights, with a median seeing of $\sim1.2\arcsec$ with standard deviation $\sim0.12\arcsec$ for all observations. Further instrument and observing details are given in MJ22.

\subsection{Mephisto} \label{sec_meph}
NGC~4395 was also observed with the 1.6~m Mephisto telescope \citep[e.g.,][]{Yuan20,Chen24} at Lijiang Observatory on 2025 April~1--2. Mephisto, the wide-field multi-channel photometric telescope, is equipped with three CCD cameras, allowing simultaneous imaging in the $u$ or $v$ (blue), $g$ or $r$ (yellow), and $i$ or $z$ (red) bands for each pointing. The observations of NGC 4395 were acquired with Mephisto in its commissioning phase. During this period, all three channels had a field of view of $44\arcmin \times 44\arcmin$. 
The red channel has a finer spatial sampling $\sim0\farcs286$ than the blue and yellow channels $\sim0\farcs429$ (\citealt{Yang24, Chen24, Du25}). For this campaign, we adopted the $ugi$ configuration to probe a potential $u$-band lag excess. High-cadence intranight monitoring was carried out with an exposure time of 150~s per frame in each band, providing $\sim$ 6 hours of continuous coverage per night. The observations were obtained during dark time under generally photometric conditions, with seeing typically in the range of $\sim0.96$--$1.22\arcsec$.

\begin{figure}[!t]
  \centering
  \includegraphics[width=\columnwidth]{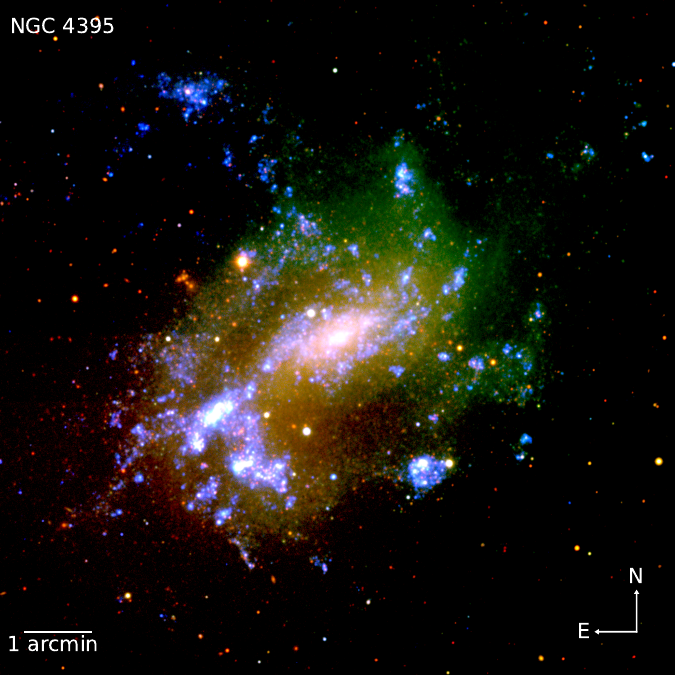}
  \caption{
  This true-color RGB image is constructed from the stacked Mephisto $u$, $g$, and $i$-band images,
  using the highest-quality 165 exposures in total and achieving a surface-brightness limit of
  26.96~mag~arcsec$^{-2}$ for $g$ band. The displayed field of view is approximately
  $10\arcmin \times 10\arcmin$. A picture of full field ($35\arcmin \times 35\arcmin$) is available 
  \href{https://doi.org/10.5281/zenodo.18149496}{here}.
  } 
  \label{fig:RGB images}
\end{figure}


\begin{figure*}
  \centering
  \begin{tabular}{c}
  \includegraphics[width=0.95\textwidth]{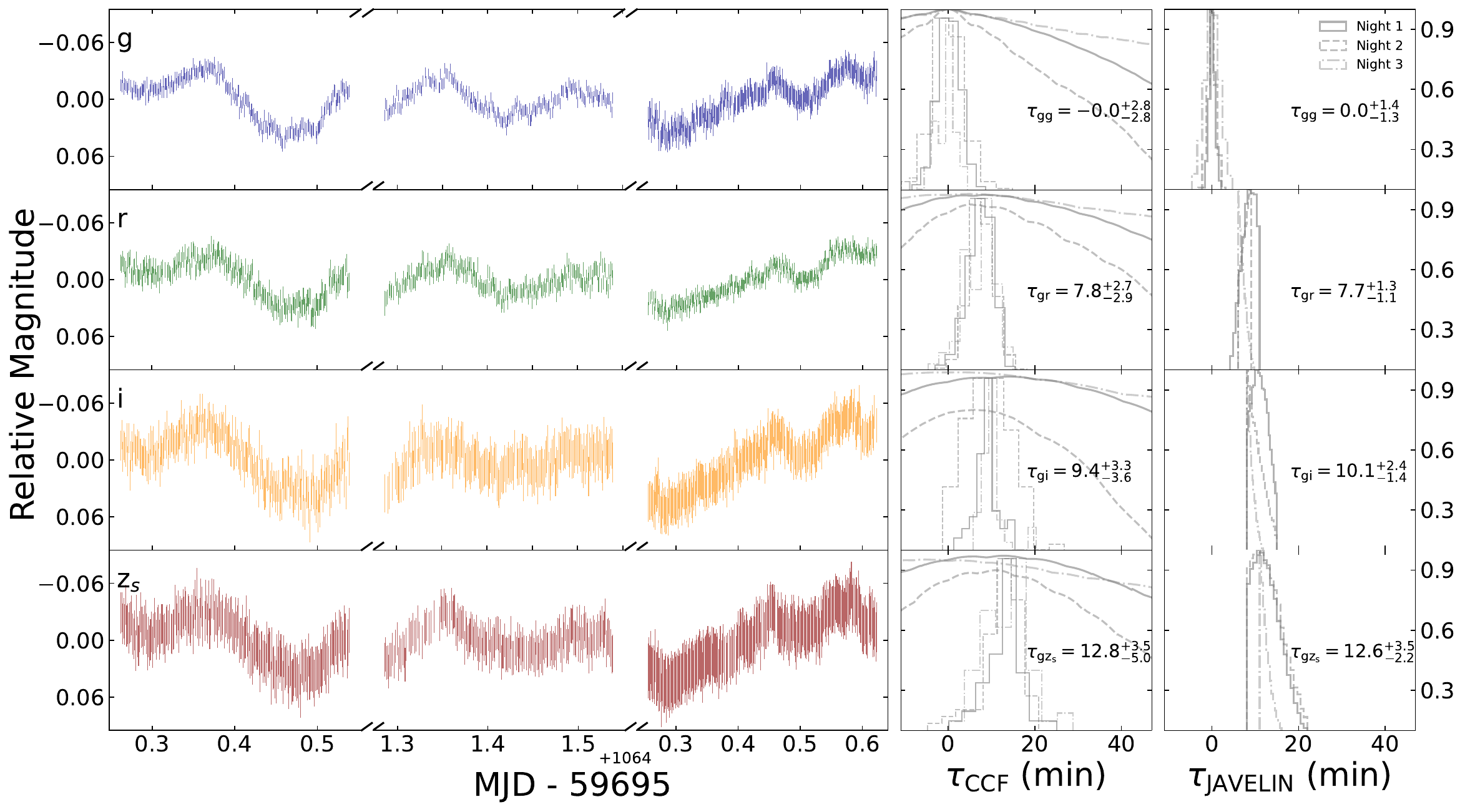}
  \end{tabular}
  \caption{Three-night multi-band FTN monitoring of NGC~4395 and the corresponding arithmetic mean lag measurements} (listed in each panel). From left to right: four-band light curves and lag measurements relative to the $g$ band using \texttt{ICCF} and \texttt{JAVELIN}. The histograms are the normalized lag posterior distributions and the curves on the top are the cross-correlation function. The data is published in its entirety in machine-readable form, the same as Figure \ref{fig:lc_Mephisto_0905}.
  \label{fig:lc_FTN_0905}
\end{figure*}

\begin{figure*}
  \centering
  \begin{tabular}{c}
  \includegraphics[width=0.95\textwidth]{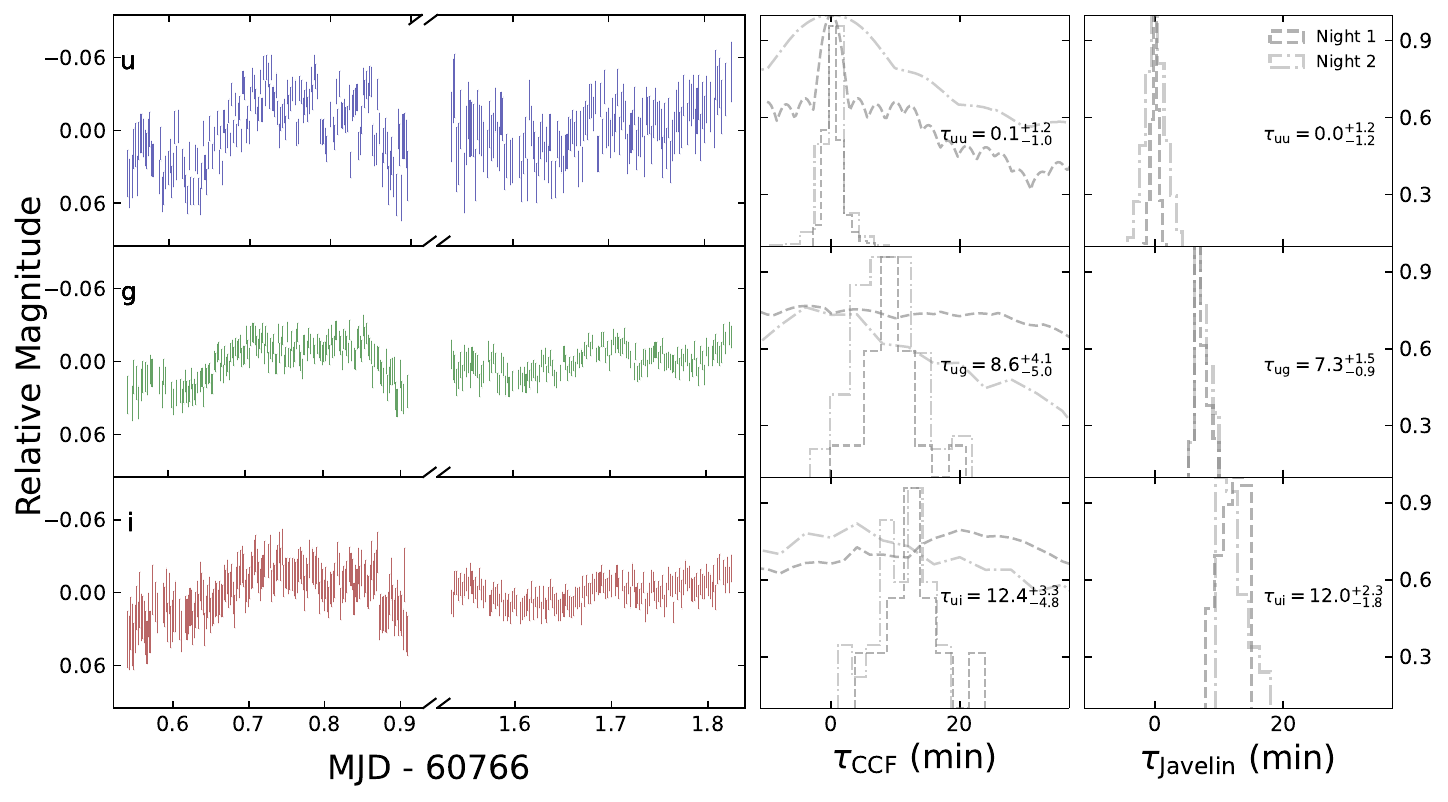}
  \end{tabular}
  \caption{Same as Figure~\ref{fig:lc_FTN_0905}, but for two-night three-band Mephisto monitoring of NGC~4395.}
  \label{fig:lc_Mephisto_0905}
\end{figure*}

\section{Data Reduction} \label{sec_reduction}
\subsection{Light curves}
All images were reduced following standard CCD processing procedures, including bad-pixel masking, bias and dark subtraction, flat-fielding, cosmic-ray removal, and astrometric calibration. The FTN data were processed using the LCO BANZAI pipeline \citep{Lang10}, while the Mephisto data were reduced with a pipeline developed by the Mephisto team \citep{Yang24,Du25}. We then employ difference-imaging techniques to extract multi-band light curves of the central nucleus, thereby minimizing contamination from the extended host galaxy. The specific steps are as follows:

\textit{Image alignment}, including both astrometric and flux alignment, was first performed to place all frames onto a common positional and flux scale, thereby enabling accurate relative photometric calibration and reducing systematic effects, e.g., arising from seeing variations and small pointing offsets between epochs.

\begin{enumerate}
\item For the FTN images, the signal-to-noise ratio (SNR) of individual exposures is sufficiently high that a single best-quality frame in each band was directly adopted as the reference for subsequent image alignment and subtraction. Given the relatively lower SNR of the Mephisto images, we selected high-quality frames with FWHM $<2\farcs5$ and ellipticity $<0.15$, and aligned and stacked them to construct the reference images in each band. Using the stacked Mephisto images, we also constructed a true-color image of NGC~4395, as shown in Figure~\ref{fig:RGB images}.

\item All single-epoch images were registered to the reference image with \texttt{SWarp} \citep{Bertin02} onto a common astrometric grid, adopting the AGN position as the alignment center.

\item Within $100\arcsec$ of the target, we selected two nearby, isolated stars whose colors and magnitudes differ from those of the target by at most 0.5~mag. These two stars were cross-matched with the ASAS-SN variable-star catalog \citep{Jayasinghe18,Jayasinghe19} to confirm that they are non-variable. The two stars were used as flux calibrators, with their measured fluxes in each exposure scaled to match those in the reference image, thereby placing all images onto a common relative photometric scale.

\end{enumerate}

\begin{table*}
\centering
\caption{Observed-frame lag measurements from FTN observations}
\label{tab:FTN lags}
\setlength{\tabcolsep}{6pt}
\begin{tabular}{l ccc ccc ccc ccc}
\hline
Band & \multicolumn{3}{c}{2022-04-26} & \multicolumn{3}{c}{2022-04-27} & \multicolumn{3}{c}{2025-03-26} & \multicolumn{3}{c}{Three nights} \\
\cline{2-13}
 & $\tau_{\rm CCF}$ & $r_{\rm max}$ & $\tau_{\rm JAVELIN}$ &
   $\tau_{\rm CCF}$ & $r_{\rm max}$ & $\tau_{\rm JAVELIN}$ &
   $\tau_{\rm CCF}$ & $r_{\rm max}$ & $\tau_{\rm JAVELIN}$ &
   $\tau_{\rm CCF}$ & $r_{\rm max}$ & $\tau_{\rm JAVELIN}$ \\
 & min &  & min &
   min &  & min &
   min &  & min &
   min &  & min \\
\hline
$g$ (4770\,\AA) &
$0.2_{-2.5}^{+2.6}$ & 1.00 & $0.0_{-0.9}^{+0.9}$ &
$-0.1_{-3.3}^{+3.3}$ & 1.00 & $0.0_{-0.1}^{+0.0}$ &
$-0.1_{-2.6}^{+1.8}$ & 1.00 & $0.1_{-2.1}^{+2.1}$ &
$-0.0_{-2.5}^{+2.9}$ & 1.00 & $0.0_{-0.6}^{+0.6}$ \\

$r$ (6215\,\AA) &
$7.6_{-2.5}^{+3.2}$ & 0.98 & $8.4_{-2.1}^{+1.7}$ &
$7.5_{-2.5}^{+3.0}$ & 0.93 & $7.7_{-1.1}^{+0.9}$ &
$8.3_{-3.9}^{+2.0}$ & 0.98 & $7.1_{-0.8}^{+1.3}$ &
$9.5_{-3.3}^{+2.3}$ & 0.92 & $9.0_{-0.6}^{+0.6}$ \\

$i$ (7545\,\AA) &
$9.1_{-2.4}^{+1.7}$ & 0.97 & $11.1_{-2.0}^{+2.2}$ &
$9.2_{-4.7}^{+4.9}$ & 0.80 & $10.2_{-1.6}^{+2.5}$ &
$10.0_{-1.8}^{+1.2}$ & 0.99 & $9.0_{-0.7}^{+1.4}$ &
$13.3_{-2.9}^{+3.3}$ & 0.96 & $13.1_{-1.8}^{+1.8}$ \\

$z_{s}$ (8700\,\AA) &
$13.1_{-3.8}^{+3.1}$ & 0.97 & $12.6_{-3.0}^{+4.0}$ &
$13.4_{-3.9}^{+3.2}$ & 0.90 & $12.9_{-3.3}^{+4.3}$ &
$12.0_{-4.3}^{+4.5}$ & 0.95 & $12.4_{-1.0}^{+2.0}$ &
$15.1_{-4.9}^{+2.3}$ & 0.93 & $15.1_{-2.9}^{+2.9}$ \\
\hline
\end{tabular}
\tablecomments{The CCF lags listed here are centroid lags, and $r_{\rm max}$ denotes the peak cross-correlation coefficient. The same definition applies to Table \ref{tab:Meplags}.}
\end{table*}

\begin{table*}
\centering
\setlength{\tabcolsep}{8pt}
\caption{Observed-frame lag measurements from Mephisto observations.}
\label{tab:Meplags}
\begin{tabular}{l ccc ccc ccc}
\hline
Band & \multicolumn{3}{c}{2025-04-01} & \multicolumn{3}{c}{2025-04-02} & \multicolumn{3}{c}{Two nights} \\
\cline{2-10}
 & $\tau_{\rm CCF}$ & $r_{\rm max}$ & $\tau_{\rm JAVELIN}$ &
   $\tau_{\rm CCF}$ & $r_{\rm max}$ & $\tau_{\rm JAVELIN}$ &
   $\tau_{\rm CCF}$ & $r_{\rm max}$ & $\tau_{\rm JAVELIN}$ \\
 & min &  & min &
   min &  & min &
   min &  & min \\
\hline
$u$ (3450\,\AA) &
$0.0^{+1.2}_{-1.1}$ & 1.00 & $0.0^{+0.7}_{-0.7}$ &
$0.0^{+1.2}_{-1.1}$ & 1.00 & $0.1^{+1.9}_{-1.9}$ &
$0.0^{+4.2}_{-4.2}$ & 1.00 & $0.0^{+0.6}_{-0.6}$ \\

$g$ (5273\,\AA) &
$8.9^{+2.4}_{-3.0}$ & 0.78 & $7.1^{+1.5}_{-0.8}$ &
$8.2^{+4.7}_{-4.7}$ & 0.76 & $7.4^{+1.4}_{-1.0}$ &
$10.5^{+4.9}_{-4.2}$ & 0.95 & $10.2^{+1.6}_{-0.9}$ \\

$i$ (8301\,\AA) &
$12.8^{+3.4}_{-4.2}$ & 0.80 & $12.1^{+2.0}_{-2.0}$ &
$11.9^{+2.5}_{-4.4}$ & 0.82 & $12.0^{+3.0}_{-1.5}$ &
$15.4^{+2.5}_{-3.3}$ & 0.96 & $14.4^{+3.0}_{-1.7}$ \\
\hline
\end{tabular}
\end{table*}

\textit{Image subtraction} was performed using the \texttt{HOTPANTS} software \citep{Becker15}. For the differencing, we adopted a convolution kernel with a half-width of 2 times the image FWHM and a substamp half-width of 2.5 times the FWHM centered on each source. After generating the difference images, forced photometry was carried out at the positions of the target and reference stars using multiple aperture sizes. An aperture radius of $\sim$3\farcs4, which minimizes the magnitude dispersion of the reference stars, was selected as the optimal aperture.

We additionally compared two alternative photometric methods, standard aperture photometry with \texttt{Photutils} \citep{Bradley16} and PSF photometry based on \texttt{PSFEx} \citep{Bertin13} and \texttt{SExtractor} \citep{Bertin96}. Among all tested approaches, the \texttt{HOTPANTS}-based measurements yield the smallest scatter for selected reference stars and are therefore adopted as our fiducial results, as shown in Figures~\ref{fig:lc_FTN_0905} and \ref{fig:lc_Mephisto_0905}.

To assess the reliability of the photometric uncertainties, we examined the light curves of the two non-variable check stars in the field. We find that the observed scatter of the check-star light curves is larger than expected from the formal photometric uncertainties, indicating that the errors are underestimated. Following common practice in time-domain photometry \citep[e.g.,][]{Zebrun01,Hartman04,Fausnaugh16}, we rescaled the photometric uncertainties using multiplicative factors derived from the reduced $\chi^2$ of the check-star light curves. The scaling factors were determined independently for each night, telescope, and filter to account for varying observing conditions, and range from $1.3$ to $4.1$. This rescaling brings the normalized residual distribution into better agreement with a unit-variance Gaussian distribution. The resulting correction factors were then applied to the uncertainties of the target light curves.

\subsection{Lag Measurements}
CRM measures time delays between variability observed in different bands, which encode the light-travel-time separation between physically distinct continuum emitting regions. Such delays are commonly estimated using either correlation-based techniques or forward modeling of the light curves. Among these, the interpolated cross-correlation function (\texttt{ICCF}; \citealt{Gaskell87,White94}) and the Bayesian modeling approach \texttt{JAVELIN} \citep{Zu11} are the two most widely adopted methods. The \texttt{ICCF} method provides a relatively model-independent estimate of the lag based on the correlation between light curves, while \texttt{JAVELIN} infers the lag by modeling the reprocessed light curve as a shifted, scaled, and smoothed response to a stochastic driving signal. Previous studies have shown that the two methods generally yield consistent lag measurements, with \texttt{JAVELIN} often reporting relatively smaller uncertainties \citep{Edelson19,LiJ19,Yu.Z19}. However, each method has its own limitations, as \texttt{ICCF} can be sensitive to sampling and noise properties, whereas \texttt{JAVELIN} relies on assumptions about the variability model and transfer function.

To ensure the robustness of our lag measurements for NGC~4395, we therefore employ both \texttt{ICCF} and \texttt{JAVELIN} as complementary and independent estimators. For both methods, we adopted a lag search window of $-70$ to $+70$ minutes with a sampling interval of 0.5 minutes. For the \texttt{ICCF} analysis, uncertainties were estimated using the standard flux randomization and random subset selection (FR/RSS) method \citep{Peterson98} with 20000 realizations, and the lag was taken as the median of the cross-correlation centroid distribution. For \texttt{JAVELIN}, the same lag window was used, with the lag determined from the median of the posterior distribution obtained via Markov Chain Monte Carlo sampling. The resulting lag measurements relative to the shortest bands in the observed frame from the FTN and Mephisto observations are listed in Tables~\ref{tab:FTN lags} and \ref{tab:Meplags}, respectively.

For each observing night, the interband lags were measured independently using the procedures described above. The lag values shown in the right panels of \ref{fig:lc_FTN_0905} and \ref{fig:lc_Mephisto_0905} correspond to the simple average of the nightly lag measurements, with the associated uncertainties estimated from the centroid-lag distributions and are defined as the 16th–84th percentile range. In addition to the per-night analysis, we also measured lags using combined light curves constructed by concatenating the data from multiple nights in chronological order. The resulting lag measurements from these combined light curves are reported as the ``Two nights'' and ``Three nights'' values in Table \ref{tab:FTN lags} and \ref{tab:Meplags}. These measurements are provided for comparison with the nightly results and to assess the effect of increased data volume on the lag determination. We adopt the ICCF lag measurements based on the combined light curves as our fiducial results for the subsequent analysis, as they fully utilize all available data and provide the most robust estimates; although the combined data show a slight increase in lag amplitudes consistent with red-noise variability, the measurements remain consistent within uncertainties and do not exhibit strong red-noise–dominated behavior.


\section{Results and Discussion} \label{sec_result and discussion}
\subsection{Continuum lag–wavelength relation} 
The resulting lag measurements are shown in Figures~\ref{fig:lc_FTN_0905} \& \ref{fig:lc_Mephisto_0905}. Generally speaking, our lag measurements are highly self-consistent across different observing nights and between the two independent datasets. Across all nights, the UV and optical light curves exhibit strong and stable correlations, indicating highly coherent continuum variability in UV/optical.

Relative to previous monitoring campaigns, our measured inter-band lags show systematic offsets in absolute amplitude: they are slightly smaller ($\sim$ 30\%) than those reported by MJ22, yet a factor of two larger than the lags inferred by MI23. To explore the origin of these systematic differences, we compared their techniques and light-curve properties. 

\begin{figure}
  \centering
   \includegraphics[width=\columnwidth]{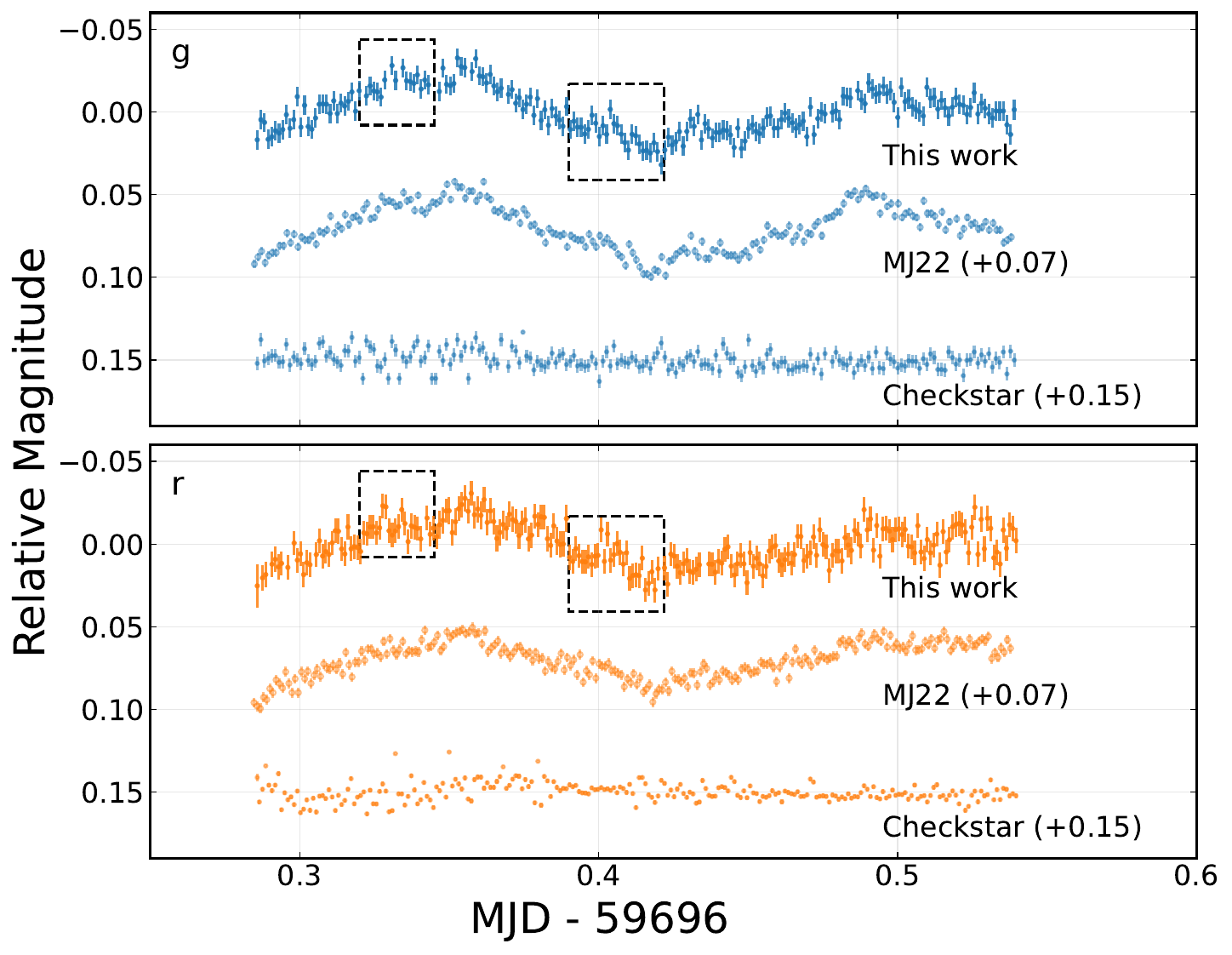}
    \caption{
Comparison of the second-night $g$- and $r$-band light curves from FTN between this work and MJ22. Both analyses use the same raw data but adopt different photometric techniques (difference imaging vs.\ standard aperture photometry). The AGN light curves are shown together with the corresponding check-star light curves derived from difference imaging. The black dashed rectangles mark two substructures identified in this work that are not clearly visible in MJ22. No clear or consistent substructures are observed in the check-star light curves at the corresponding epochs; localized similarities may appear at other epochs.}
\label{fig:compare with Montano22}
\end{figure}

For the first two observing nights, our analysis is based on the same FTN datasets used by MJ22, indicating that these differences arise from the photometric methodologies rather than from the underlying observations. Standard aperture photometry, as employed by MJ22, is straightforward and widely used, but does not explicitly account for seeing variations. Difference-imaging techniques, as adopted in this work, explicitly model and remove seeing-dependent components and static background emission, thereby more effectively isolating intrinsic nuclear variability, although the increased complexity of the procedure may introduce additional uncertainties and lead to larger dispersion in the reconstructed light curves. As shown in Figure~\ref{fig:compare with Montano22}, thanks to the use of difference-imaging photometry, our light curves exhibit slightly larger variability amplitudes and more pronounced short-term substructures than those reported by MJ22. These substructures appear at consistent temporal locations across different bands, while showing no clear correlation with seeing variations. At the epochs where the prominent substructures are identified in the target light curves, no clear counterparts are seen in the check-star light curves. At other epochs, localized similarities between the target and check-star light curves are observed, likely reflecting residual systematic effects. In particular, substructures associated with obvious lags may trace X-ray reprocessing in the accretion disk, whereas those without detectable lags could be related to localized disk processes, such as hot spots, magnetohydrodynamic turbulence, or magnetic reconnection events in the accretion flow \citep{Dexter11,Nowak12,Jiang19,Sun20}. Nevertheless, we cannot completely rule out the possibility that a fraction of the observed substructures is influenced by residual photometric uncertainties.

On the other hand, the seeing-decorrelation method adopted by MI23 provides an alternative strategy for mitigating atmospheric effects. Their observations were obtained with a significantly higher temporal cadence (3~s exposures in $g$, $r$, $i$, and $z$), which in principle improves the sampling of rapid variability. However, the total effective monitoring duration in MI23 is relatively short, at only $\sim$2~hours, much shorter than the typical baseline of our campaigns. As a result, the observed variability amplitude within such a limited time span is relatively small, which reduces the contrast of variability features in the light curves. This, in turn, weakens the ability to robustly measure interband lags and can lead to a systematic underestimation of the time delays.

To quantify the impact of variability amplitude and the presence of nearly zero-lag substructures, we performed a suite of dual-band light-curve simulations. In these simulations, the intrinsic continuum variability was modeled as a damped random walk with a fixed input inter-band lag of 20~min. Pairs of light curves were generated with a cadence of 150~s over a total duration of 25,000~s ($\approx 6.9$~hr) and were analyzed using the same ICCF procedure applied to the observational data (see Appendix~\ref{sec_appendix} for full details). Figure~\ref{fig:amplitude simulation result} summarizes the results. We find that (1) decreasing the intrinsic variability amplitude leads to systematically smaller recovered lags and larger scatter relative to the input lags, and (2) adding nearly zero-lag substructures suppresses the recovered lag, with the effect becoming stronger as either their number or their characteristic timescale increases. These results demonstrate that differences in variability amplitude and light-curve substructure alone can naturally produce the systematic offsets, without requiring intrinsic changes in the underlying lag–wavelength relation.

When the light curves from the two Mephisto nights are combined and the three FTN nights are combined separately, the inter-band lags measured from these joint data sets are slightly larger than those from individual nights, but remain statistically consistent within the uncertainties. Such behavior is commonly associated with red-noise AGN variability \citep{Press78,Edelson24}, in which longer-term fluctuations dominate the variability power and, when convolved with a broad and asymmetric transfer function, preferentially weight its extended, long-lag tail. In these cases, slowly varying components may contribute significantly to the measured lags, leading to an apparent increase in lag amplitude with longer monitoring baselines. However, in our data the combined-light-curve lag measurements do not show evidence that the lags are being driven by such long-timescale, slowly varying components: the lag–wavelength relation and its overall amplitude remain stable when going from single-night to combined-night analyses. This indicates that red-noise–dominated long-term variations (e.g., DC components) do not play a major role in shaping our measurements. We therefore adopt the combined measurements as our fiducial results, as they make use of all available data and provide the most robust estimate of the lag–wavelength relation.

Based on the fiducial lag measurements described above, we modeled the observed continuum time lags using a power-law lag--wavelength relation $\tau(\lambda) = \tau_0 \left[ (\lambda / \lambda_0)^\beta - 1 \right]$, where the reference wavelength is $\lambda_0 = 4770~\text{\AA}$. By fixing the slope to $\beta = 4/3$, as predicted by the SSD model, we find that both the FTN $griz$ measurements and the combined FTN and Mephisto dataset are well described by the same lag--wavelength relation, with best-fitting normalization $\tau_0 = 14.07^{+1.90}_{-1.88}~\mathrm{min}$. Allowing the slope $\beta$ to vary freely yields a best-fitting value of 
$\beta = 0.87^{+0.70}_{-0.30}$ for the FTN data alone, which remains consistent with the canonical thin-disk prediction of $\beta = 4/3$ within the uncertainties, and the joint fit including Mephisto data gives a statistically consistent result. 
Overall, the lag--wavelength relation closely follows the expected power-law scaling $\tau \propto \lambda^{4/3}$, and we find no evidence for a flattening of the relation at longer wavelengths than $\sim8000~\text{\AA}$, suggesting that our observations do not yet probe the outermost regions of the accretion disk and do not reveal compelling evidence for a bowl-shaped torus geometry \citep{Starkey22,Edelson24}.

\begin{figure*}
  \centering
  \includegraphics[width=0.8\textwidth]{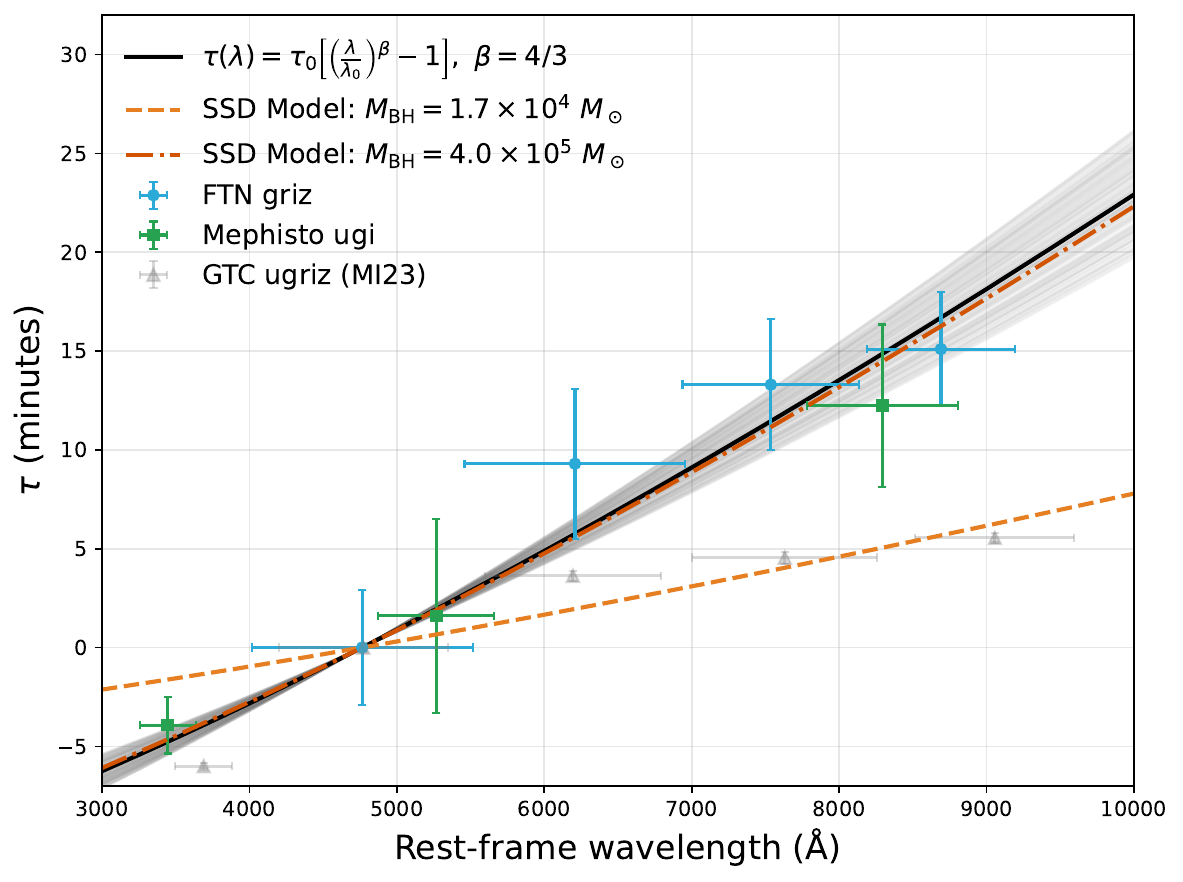}
    \caption{Lag--wavelength relation in the rest frame of NGC~4395. Blue points show the \texttt{ICCF} lags measured from the combined three nights of FTN data. Green points show the Mephisto lags converted to relative delays with respect to $4770\,\mathrm{\AA}$ (the FTN $g$ band).This conversion is performed independently of the FTN fit and therefore does not introduce any bias in the comparison between the two datasets. Gray points denote the GTC measurements reported by MI23. The black curve represents the best-fitting lag--wavelength relation obtained from the FTN $g r i z$ data alone, assuming a fixed slope of $\beta=4/3$.  The gray curves illustrate the $1\sigma$ uncertainty estimated from the MCMC analysis. The orange dashed and dotted curves show the predictions of the standard thin-disk model for black hole masses of $M_{\rm BH}=1.7\times10^{4}\,M_\odot$ and $4\times10^{5}\,M_\odot$, respectively.} 
\label{fig:all_lamda_tao_gz}
\end{figure*}

Regarding the inferred disk size, the lag measurements in Figure~\ref{fig:all_lamda_tao_gz} are broadly consistent with the SSD expectation for a black hole mass of order $10^{5}\,M_{\odot}$, as independently indicated by gas dynamical constraints \citep{denBrok15,Brum19} and \ion{C}{4} RM \citep{Peterson05}. However, it would instead imply a disk size larger by a factor of $\sim$2 if adopting a lower mass of $\sim10^{4}\,M_{\odot}$ based on H$\beta$ RM and CRM \citep{Woo19,Wang23}. This disk size also appears more consistent with MI23. Considering the DC contribution is not important here (see spectral decomposition results in \S \ref{sec:spec_decom}), this suggests that other mechanisms may play a role in driving AGN variability. Overall, the lag-inferred continuum-emitting size does not significantly deviate from the expectations of an SSD plus X-ray reprocessing scenario. Whether it follows the SSD prediction precisely or allows room for additional variability mechanisms depends on the measurement uncertainties in both the black hole mass and the continuum lags.

\begin{figure*}
  \centering
  \includegraphics[width=1.0\textwidth]{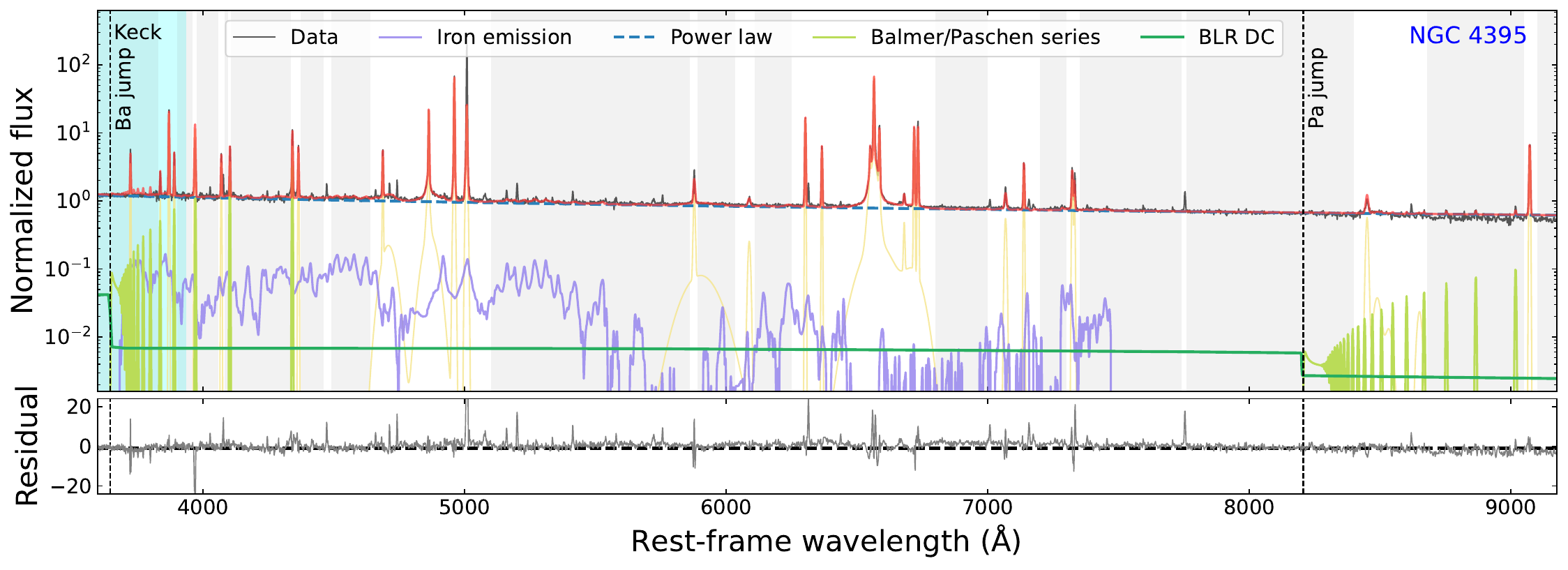}
    \caption{Spectral decomposition of the optical spectrum of NGC~4395 based on combined Keck and SDSS DR18 data. The red curve shows the best-fitting model, which includes the following non-zero components: a power-law accretion-disk continuum (blue dashed), Fe~II emission (purple), the Balmer and Paschen line series (light green), diffuse continuum emission from the BLR (dark green), and emission lines (yellow). Gray shaded regions indicate the continuum fitting windows used in the fit, excluding strong emission lines, while the cyan shaded region marks the wavelength range covered by the Keck spectrum. The vertical black dashed lines mark the locations of the Balmer jump (3646~\AA) and the Paschen jump (8204~\AA). No significant host-galaxy component is recovered, as the host is spatially extended and most of its emission falls outside the SDSS aperture.}
\label{fig:NGC4395-DC}
\end{figure*}

\subsection{U/u-band lag excess}\label{sec:spec_decom}
Lag excess around the Balmer break at 3646\AA\ is a key feature to evaluate the DC contribution to the continuum. Based on the two-night Mephisto monitoring, we have not detected any significant $u$-band lag excess. This result is consistent with the high-cadence multi-band GTC observations MI23, and suggests that DC emission does not dominate the optical continuum lags at least during the recent epochs.

To further quantify the fractional contribution of DC emission, we performed a spectral decomposition of the SDSS spectrum obtained in 2006 March\footnote{Since the SDSS spectrum only extends down to $\sim$3800~\AA, we extrapolate to $\sim$3550~\AA\ to cover the Balmer break using the Keck/LRIS spectrum obtained on 2019 March 3 and April 2 with the Low Resolution Imaging Spectrometer on the Keck~I 10-m telescope and digitized from \citet{Cho21}. Because the SDSS and Keck spectra were obtained at different epochs separated by more than a decade, variability in the AGN continuum and emission-line strengths may introduce additional systematic uncertainties in the spectral decomposition.} using \texttt{PyQSOFit} \citep{Guo18}, following the methodology of \citet{Guo22a} and \citet{Gonzalez-Buitrago25}. We briefly outline the basics below and refer the reader to references for details. A key aspect of this approach is the explicit inclusion of a DC component together with high-order Balmer/Paschen series, which jointly compensate for the intrinsic Balmer/Paschen-edge discontinuity of the DC, thereby reconciling the observed generally flat continuum shape. The model consists of a power-law continuum, host-galaxy emission, Fe \textsc{II} emission, a BLR DC component obtained from \citet{Korista19}, high-order Balmer/Paschen line series, and Gaussian profiles for other emission lines. In the fitting, the DC component is assumed to share the same Doppler velocity broadening as the Balmer line series.

Figure~\ref{fig:NGC4395-DC} shows that the fractional contribution of the DC component is $\sim$2.7\% around the Balmer break relative to the total observed flux, while its contribution in the Paschen jump region is negligible ($\sim$0.7\%). Assuming that the spectral trend extends smoothly across the full bandpass, this corresponds to a DC contribution of $\sim$1.1\% when convolved with a realistic $u$-band filter. The statistical uncertainties ($\sim$0.5\%) estimated from Monte Carlo realizations are small compared to the dominant systematic uncertainties associated with model degeneracies. To assess the robustness of the inferred DC strength, we explored alternative continuum prescriptions (e.g., a SSD disk or a broken power law) and also performed local fits restricted to the Balmer break region ($3500$–$4500$~\AA). All approaches yield consistent DC fractions in the range of $\sim2.7–13.4$\% around Balmer break (or $\sim0.7-3.4$\% around Paschen break), which is significantly smaller than normal Seyfert 1 AGNs with a fraction of $10-50$\% at Paschen break \citep{Guo22a}.

As expected, spectral decomposition results also indicate that the DC contribution in NGC~4395 is minimal. Previous studies have shown that the DC component depends sensitively on the ionizing luminosity: it is often undetectable during low-luminosity states but becomes significant as the source brightens \citep[e.g., Mrk~110,][]{Vincentelli21,Vincentelli23}.  Given its intrinsically low luminosity, NGC~4395 is therefore expected to produce only a weak DC component, which may help stabilize the measured continuum lags over multi-year timescales and yields lag-inferred continuum sizes that are closer to the predictions of the SSD model.

\subsection{Multi-year lag stability} 

Based on our five-night monitoring campaigns, we find that the optical continuum lags in NGC~4395 exhibit remarkable stability over month-to-year timescales. We propose that the following factors may play important roles:

{\it Intrinsically stable driving light curves} Long-term X-ray studies indicate that the hard X-ray (2–10~keV) luminosity of NGC~4395 has remained broadly consistent in past decades, with the flux varying by no more than a factor of $\sim$2 \citep{ONeill06,Kammoun19}. Likewise, contemporaneous ZTF optical monitoring reveals only mild long-term variability from 2022 to 2025, with the optical flux density declining gradually from $\sim1.65$~mJy to $\sim1.15$~mJy \citep{Sun25}. This further indicates that the DC contribution is negligible, since a DC-dominated scenario following an $R$–$L$ relation with $\beta=0.5$ \citep{Netzer22,Guo22} would predict a measurable lag decrease of $\sim$17\% for the observed $\sim$30\% luminosity decline, which is not observed.

{\it Stable disk-corona geometry} Given its low luminosity and inferred average Eddington ratio of $\sim 0.06$ \citep{Kammoun19}, this source is comparable to other nearby Seyfert galaxies such as NGC~4151, NGC~4593, and NGC~5548. NGC~4395 likely lies near the boundary between the standard thin-disk regime and a radiatively inefficient accretion flow (ADAF; \citealt{Narayan95}). A further decline in luminosity could push the source into an ADAF state, characterized by a hot, optically thin, geometrically thick flow with low radiative efficiency, which would suppress UV/optical emission and significantly alter reprocessing signatures. The weak long-term variability in NGC~4395 instead indicates that the source has remained in a standard accretion mode during our observations, implying a stable disk–corona geometry and, consequently, stable continuum lags.

{\it X-ray loudness helps stabilize continuum lags} The relatively high X-ray–to–optical luminosity ratio of NGC~4395 favors a variability regime dominated by X-ray reprocessing. Like other low-luminosity AGNs, NGC~4395 is X-ray loud, with $\alpha_{\rm ox}\approx-1.03$\footnote{Quantitatively, adopting $L_{5100}=5.75\times10^{39}\,{\rm erg\,s^{-1}}$ \citep{Cho20} and $L_{2-10\,{\rm keV}}\simeq1.5\times10^{40}\,{\rm erg\,s^{-1}}$ \citep{Kammoun19}, we estimate $\alpha_{\rm ox}\approx-1.03$, assuming $\alpha_{\nu}\simeq-0.5$ and $\Gamma\simeq1.6$ \citep{Cai24}.}, which is $\sim$3$\times$ higher than that of typical Seyfert~1 galaxies ($\alpha_{\rm ox}\approx-1.2$) and $\sim$10$\times$ higher than that of luminous quasars ($\alpha_{\rm ox}\approx-1.4$) \citep{Ho09}. As a result, the UV/optical continuum may primarily respond to variations in a compact central X-ray source, potentially producing tightly correlated multi-wavelength light curves. This expectation is consistent with the generally strong X-ray–UV/optical correlations previously reported for this source \citep{Desroches06,ONeill06,Cameron12}. In such a regime, reprocessing-driven variability is expected to dominate over other mechanisms (e.g., localized disk inhomogeneities or transient hot spots), naturally leading to relatively stable continuum lags.

\section{Conclusions}\label{sec_summary}

We performed high-cadence, multi-band monitoring of the prototypical IMBH hosted by the Seyfert~1 dwarf galaxy NGC~4395 using FTN and Mephisto telescopes. Our main findings can be summarized as follows:

\begin{enumerate}

\item The UV/optical light curves are strongly correlated, with inter-band lags increasing monotonically with wavelength, consistent with the slope expected by SSD model. No obvious lag flattening is observed at $\gtrsim$8000~\AA.

\item The DC contribution to the optical continuum is negligible, as indicated by the absence of a $u$-band lag excess and independently supported by spectral decomposition including an explicit DC component.

\item In general, lag-inferred infrared continuum–emitting sizes do not deviate from SSD expectations by more than a factor of two. Determining whether they are fully consistent with the theoretical prediction requires more accurate measurements of both the black hole mass and the continuum lags.

\item The remarkable multi-year stability of the optical continuum lags may be related to its intrinsically modest long-term variability, a stable disk–corona geometry and an unusually high X-ray-to-optical luminosity ratio leading X-ray reprocessing dominates the UV/optical variability over other potential variability mechanisms.

\end{enumerate}

The future facility Gemini/SCORPIO, a simultaneous 8-channel, optical to infrared camera and spectrograph\footnote{\url{https://www.gemini.edu/instrumentation/future-instruments/scorpio}} \citep{Diaz22}, will enable a major advance in CRM field. Its strictly simultaneous, wide wavelength coverage is ideal for measuring short, wavelength-dependent continuum lags while minimizing systematics from non-simultaneous observations, providing stringent constraints on accretion-disk structure and diffuse continuum emission models.

\vspace{0.2in}

Mephisto is developed at and operated by the South-Western Institute for Astronomy Research of Yunnan University (SWIFAR-YNU), funded by the ``Yunnan University Development Plan for World-Class University'' and the ``Yunnan University Development Plan for World-Class Astronomy Discipline''. The authors acknowledge support from the ``Science \& Technology Champion Project'' (202005AB160002) and from two ``Team Projects'' --- the ``Innovation Team'' (202105AE160021) and the ``Top Team'' (202305AT350002), all funded by the ``Yunnan Revitalization Talent Support Program''. This work is also supported by the National Key Research and Development Program of China (2024YFA1611603) and the ``Yunnan Provincial Key Laboratory of Survey Science'' with project No.~202449CE340002. This work is also supported by the National Key R\&D Program of China (No.~2023YFA1607903, 2022YFF0503402). H.~X.~G. acknowledges support from the National Natural Science Foundation of China (NSFC, No.~12473018, 12522304) and the Overseas Center Platform Projects, CAS, No.~178GJHZ2023184MI.

\facility{LCOGT, Mephisto}
\software{AstroPy \citep{Astropy18}}

\appendix 

\section{Example Simulated Light Curves}
\label{sec_appendix}

To investigate how variability amplitude and short-term substructures affect multi-band lag measurements, we performed two sets of dual-band light-curve simulations using the \texttt{pyLCSIM} package \citep{Campana17}, following the methodology of \citet{Guo17}. The intrinsic continuum variability was modeled as a damped random walk \citep[DRW;][]{Kelly09}. In all simulations, pairs of light curves were generated with a fixed input inter-band lag of 20~min, sampled with a cadence of 150~s over a total duration of 25{,}000~s ($\approx 6.9$~hr). Gaussian photometric noise was added to match the typical cadence and uncertainty levels of the FTN observations.

In the first set of simulations, we examined the impact of variability amplitude. All parameters, including the intrinsic lag, cadence, and noise properties, were held fixed, while the peak-to-peak variability amplitude was varied between approximately 0.05 and 0.10~mag. For each amplitude, 1000 independent pairs of light curves were generated and analyzed using the same ICCF procedure applied to the observational data. Left panel of Figure~\ref{fig:simulation hist result} shows that reducing the variability amplitude leads to a systematic decrease in the recovered mean lag by approximately 25\%, accompanied by a broadening of the lag distribution by a factor of $\sim$2. The lag uncertainties are estimated from these centroid-lag distributions and are defined as the 16th--84th percentile range.

In the second set of simulations, we explored the effect of short-term substructures on lag recovery. On top of the underlying DRW variability, one or two localized substructures were introduced simultaneously in both bands, carrying no intrinsic inter-band lag. These substructures were modeled as Gaussian perturbations, with their characteristic widths varied to probe features with different durations. Figure~\ref{fig:simulation hist result} (right panel) shows that the recovered lag distributions are progressively suppressed as the number of zero-lag substructures increases and as their temporal widths become broader.

Figure~\ref{fig:simulation lc amplitude result} shows a representative realization from the variability-amplitude simulations, illustrating how reduced variability amplitude manifests as a weaker apparent lag in individual light curves. Figure~\ref{fig:simulation lc result} shows a representative realization from the substructure simulations, illustrating how zero-lag substructures can locally bias the recovered lag in individual cases.

\setcounter{figure}{0}
\renewcommand{\thefigure}{S\arabic{figure}}
\begin{figure*}
  \centering
   \includegraphics[width=0.85\textwidth]{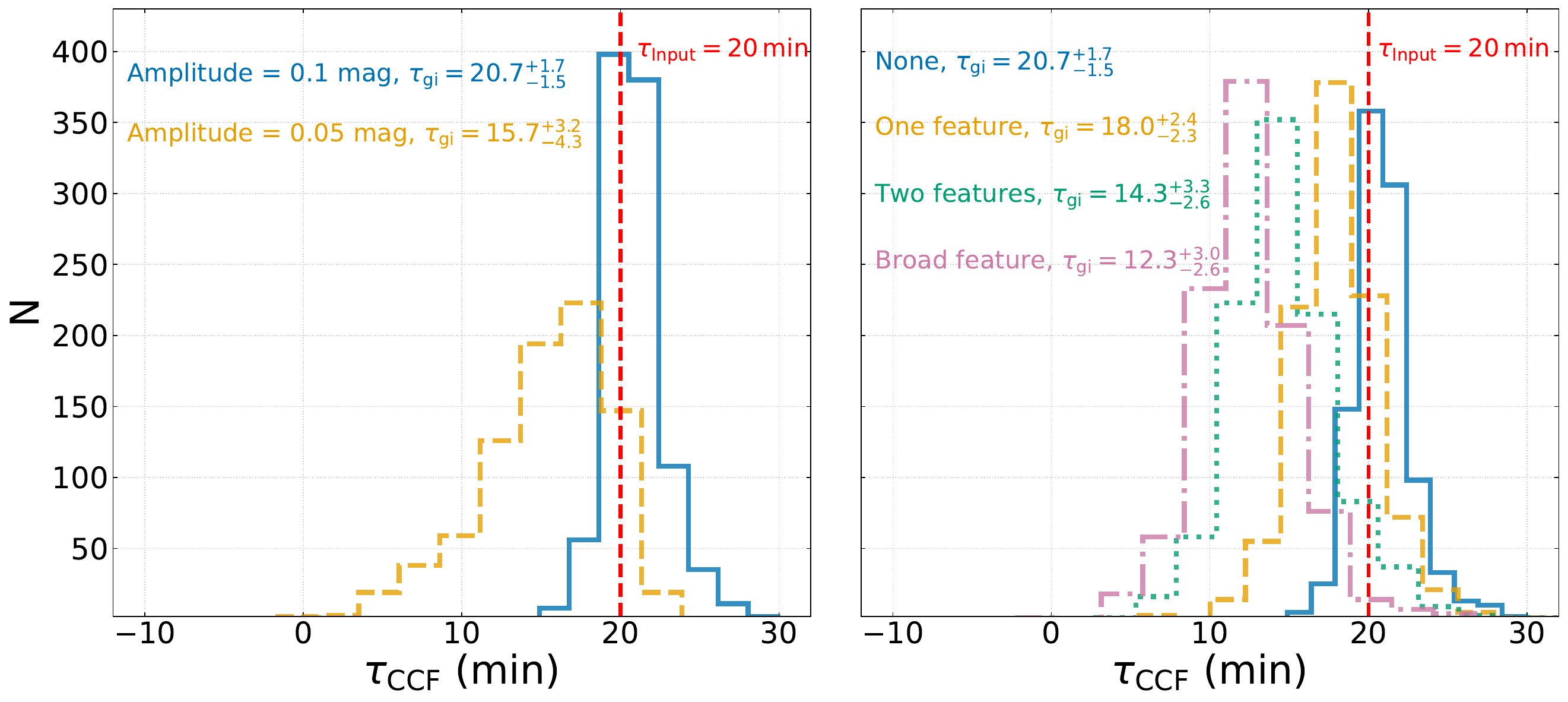}
    \caption{Lag distributions from simulated dual-band light curves. Left panel: Effect of variability amplitude on lag measurements. The blue histogram shows 1000 simulations with a variability amplitude of 0.10 mag and an input lag of 20 min (red dashed line), while the yellow histogram shows simulations with a reduced variability amplitude of 0.05 mag. Right panel: Effect of zero-lag substructures on lag measurements. All simulations share the same variability amplitude and input lag. The yellow, green, and pink histograms correspond to simulations including one, two, and two broadened zero-lag substructures, respectively. See also Appendix~\ref{sec_appendix}.}
\label{fig:simulation hist result}
\end{figure*}

\begin{figure*}[htbp]
  \centering
   \includegraphics[width=0.6\textwidth]{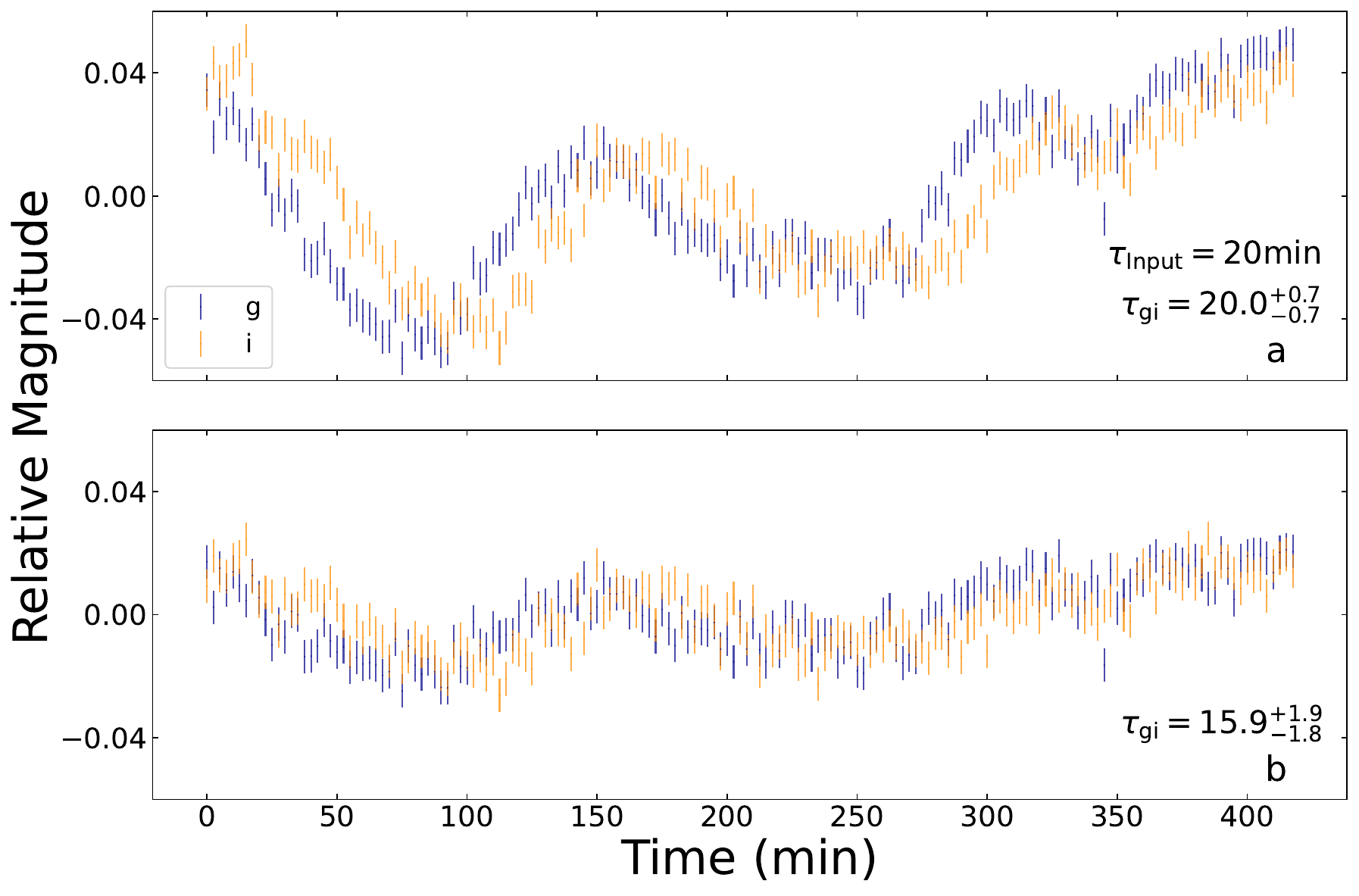}
    \caption{A representative realization from the variability-amplitude simulations. Panel (a) shows the original dual-band light curves with a variability amplitude of 0.10~mag and an input lag of 20~min, while panel (b) shows the same realization with the variability amplitude reduced to 0.05~mag, illustrating the suppression of the recovered lag.}
\label{fig:simulation lc amplitude result}
\end{figure*}

\begin{figure*}[htbp]
  \centering
   \includegraphics[width=0.6\textwidth]{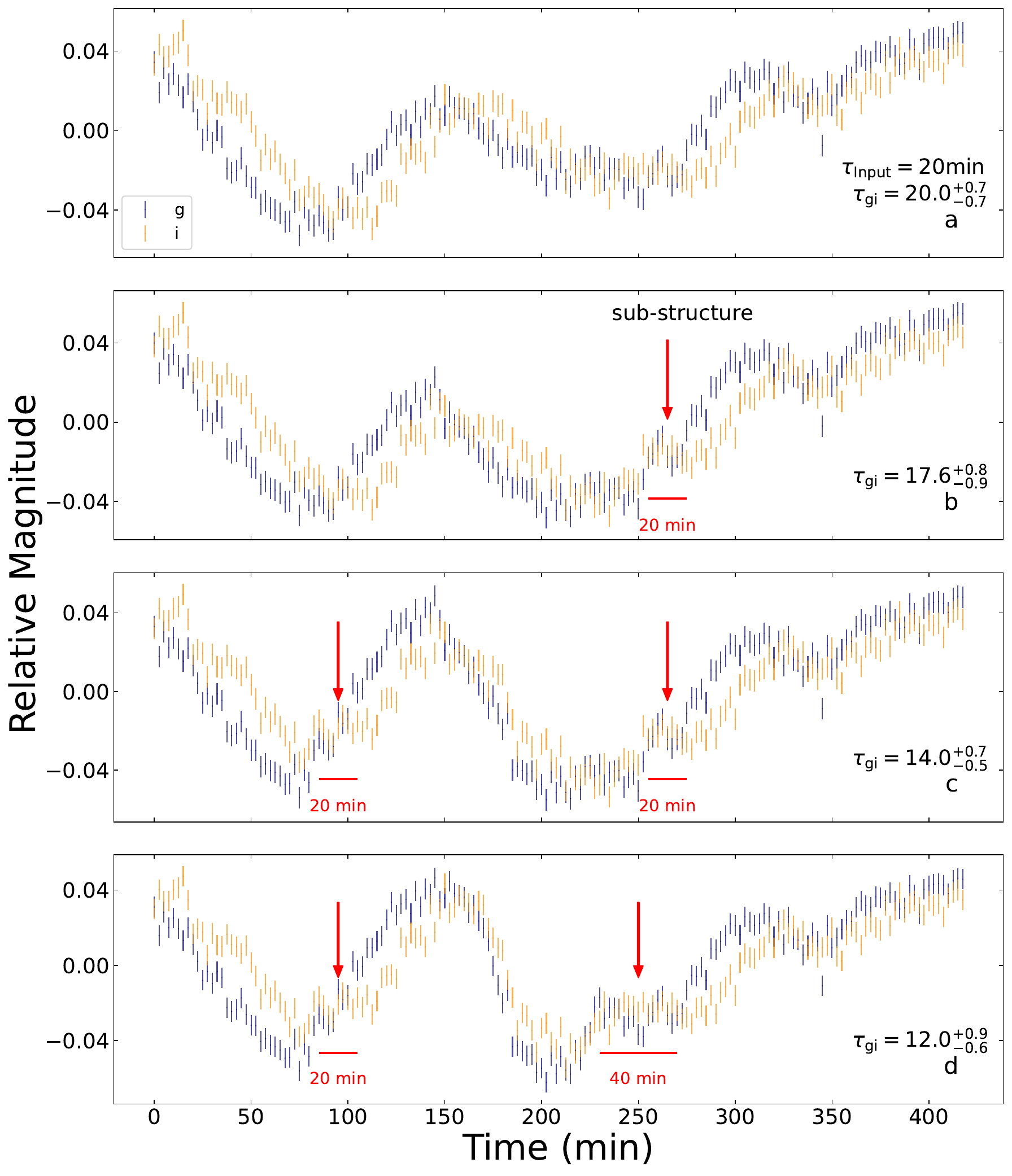}
    \caption{One representative realization from the substructure simulations. 
Panels (a)--(d) display the $g$- and $i$-band light curves and the corresponding lag measurement for different substructure configurations.
Panel (a) shows a simulated light curve without any substructure, with a variability amplitude of 0.10~mag and an input lag of 20~min.
Panel (b) shows the same light curve with one zero-lag substructure added, with an amplitude of 0.03~mag and a width of 20~min.
Panel (c) shows the case with two lag-free substructures, each having an amplitude of 0.03~mag and a width of 20~min.
Panel (d) is similar to panel (c), but with the width of one substructure increased to 40~min.
Red arrows and horizontal bars mark the locations and temporal widths of the substructures.
}
\label{fig:simulation lc result}
\end{figure*}

\bibliography{sample701}{}
\bibliographystyle{aasjournalv7}



\end{document}